\ifx\mnmacrosloaded\undefined 
\input mn\fi
\input psfig.sty


\def\rej{RE~J1034+396}

\def\eg{{e.g.~}}
\def\ie{{i.e.~}}
\def\etal{et~al.~}
\def\eV{e\kern-.15em V}                 
\def\keV{ke\kern-.15em V}                

\def\lya{Ly$\alpha$}
\def\Ha{H$\alpha$}
\def\Hb{H$\beta$}

\def\ciii{C{\sc iii}]}
\def\mgii{Mg{\sc ii}}
\def\heii{He{\sc ii}}
\def\heiifull{He{\sc ii}$\lambda$1640}

\def\oiiiab{[O{\sc iii}]$\lambda\lambda$4959,5007}
\def\ciiifull{C{\sc iii}]$\lambda$1909}

\def\civ{C{\sc iv}$\lambda$1549}
\def\mgiifull{Mg{\sc ii}$\lambda$2798}

\def\fexi{[Fe{\sc xi}]$\lambda$2649}
\def\fexiopt{[Fe{\sc xi}]$\lambda$7892}

\def\ktbb{$kT_{\rm BB}$}

\def\ax{$\alpha_{\rm x}$}

\def\auv{$\alpha_{\rm uv}$}

\def\kms{km~s$^{-1}$}

%

\newif\ifAMStwofonts

\ifCUPmtplainloaded \else
  \NewTextAlphabet{textbfit} {cmbxti10} {}
  \NewTextAlphabet{textbfss} {cmssbx10} {}
  \NewMathAlphabet{mathbfit} {cmbxti10} {} 
  \NewMathAlphabet{mathbfss} {cmssbx10} {} 
  \ifAMStwofonts
    \NewSymbolFont{upmath} {eurm10}
    \NewSymbolFont{AMSa} {msam10}
    \NewMathSymbol{\upi}     {0}{upmath}{19}
    \NewMathSymbol{\umu}     {0}{upmath}{16}
    \NewMathSymbol{\upartial}{0}{upmath}{40}
    \NewMathSymbol{\leqslant}{3}{AMSa}{36}
    \NewMathSymbol{\geqslant}{3}{AMSa}{3E}

  \else
    \def\umu{\mu}
    \def\upi{\pi}
    \def\upartial{\partial}
  \fi
\fi


\pageoffset{-2.5pc}{0pc}

\loadboldmathnames



\pagerange{000--000}    
\pubyear{0000}
\volume{000}

\begintopmatter  

\title{The UV spectrum of the narrow-line Seyfert~1 galaxy, RE~J1034+396}
\author{E. M. Puchnarewicz$^1$, K.~O.~Mason$^1$ and A. Siemiginowska$^2$}

\affiliation{$^1$Mullard Space Science  Laboratory, University College London,
Holmbury St. Mary, Dorking, Surrey RH5 6NT, UK.}
\affiliation{$^2$Harvard-Smithsonian Center for Astrophysics, 60 Garden Street,
Cambridge, MA 02138, USA.}

\shortauthor{E. M. Puchnarewicz \etal}
\shorttitle{UV spectrum of \rej}



\abstract {\rej\ has one of the hottest big blue bumps of any Seyfert 1
(\ktbb$\sim$120~\eV) and thus provides a valuable insight into the physics in
the nuclei of active galaxies. In this paper, we present UV spectroscopy of
\rej, taken with the Faint Object Spectrograph on the {\sl Hubble Space
Telescope}. With a spectral resolution of $\sim$1-2~\AA\ FWHM and a typical
signal-to-noise ratio of $\sim$15 per diode, this is one of the first detailed
UV spectra of an object in the narrow-line Seyfert 1 (NLS1) class. The spectrum
probes the physics and kinematics of the high-ionization and coronal line gas,
and the strength and form of the big blue bump component in the UV.

We detect many emission lines, including \lya, \civ, \heiifull, \ciiifull\ and
\mgiifull. We also identify a feature at 2647~\AA\ (in the rest-frame) with
highly-ionized iron (\fexi); a line of the same species (\fexiopt) has also
been seen in the optical spectrum. The velocity widths of the UV lines are
relatively narrow (FWHM$<$2000~\kms) although \civ\ appears to have a broad
underlying component with a FWHM typical of quasars ($\sim$5500~\kms). The FWHM
are similar to those of the optical lines, which suggests that {\sl all} line
emission in \rej, \ie including that of high- and low-ionization species {\sl
and} the forbidden lines, may be dominated by an intermediate-velocity
(FWHM$\sim$1000~\kms), intermediate-density (log $N_{\rm e}\sim7.5$ cm$^{-3}$)
region of gas. The slope of the UV continuum (\auv$\sim$0.9) is soft (\ie red)
relative to quasars and the UV-to-soft X-ray flux ratio is unusually low (the
0.2~\keV/1200~\AA\ flux ratio is 1/200), implying that the big blue bump
component is very weak in the UV. The present epoch UV to soft X-ray continuum
is consistent with earlier data, demonstrating that this extreme big blue bump
component is also very stable, unlike many other NLS1s which show extreme
patterns of variability.

}


\maketitle  

\section{Introduction}

\rej\ (also known as Zw~212.025) is a narrow-line Seyfert 1 galaxy (NLS1) and a
rare EUV-selected AGN (Pounds \etal 1993; Mason \etal 1995). Observations of
the 0.1-2.4~\keV\ spectrum with the  {\sl ROSAT} Position Sensitive
Proportional Counter (PSPC), showed that it has an unusually high temperature
soft X-ray component, with a high-energy turnover at around 0.4~\keV\
(Puchnarewicz \etal 1995; Pounds, Done \& Osborne 1995). An attempt was made by
Puchnarewicz \etal (1995) to measure the UV spectrum with {\sl IUE}, but the
source was found to be very faint in the UV relative to the X-rays, suggesting
the lack of any big blue bump (BBB; \eg Edelson \&\ Malkan 1986; Elvis \etal
1994) emission down to $\sim$1200~\AA. This was a surprising result,
considering the extreme EUV nature of the source, and suggested that the `big
bump' (\ie a single component which incorporates the optical/UV BBB and the
soft X-ray excess) is so hot that it is shifted out of the UV completely but
dominates in the EUV and soft X-rays.

It has now been well-established that there is a strong link between  the slope
of the soft X-ray spectrum, \ax, and the velocity of the  low-ionization
line-emitting clouds in  Seyfert 1s and quasars (\eg Puchnarewicz \etal 1992;
Boller, Brandt \& Fink 1996; Laor \etal 1997). As a NLS1 with an ultrasoft
X-ray spectrum, \rej\ is entirely consistent with this behaviour, lying at one
extreme of the relationship between \Hb\ FWHM and \ax. 

However, very little is known of the effects, if any, of the unusual continuum
shape  on the high-ionization line region. Reverberation mapping models of
`normal' Seyfert 1s, \eg NGC~5548 (\eg Clavel \etal 1991; Krolik \etal 1991),
suggest a large degree of stratification in the broad line region (BLR), \ie
the high-ionization lines (HILs; FWHM$\sim$5000-6000~\kms) are produced
relatively close to the black hole, while the low-ionization lines (LILs) are
formed light-days to light-weeks away from the HILs (FWHM$\sim$3000~\kms). A
key issue in the study of NLS1s is whether the characteristics of the HILs in
these objects are also related to the soft X-ray spectrum and form at
relatively low velocities, or whether the line-of-sight velocities of the more
highly ionized line-emitting gases are independent of \ax.

Another important issue is the nature of the UV continuum itself. The lack of
any BBB emission in the UV as suggested by the {\sl IUE} data, places a very
tight constraint on accretion disc (AD) models of the BBB. The softness of the
optical/UV continuum and lack of any rise towards the blue, implies a very hot
AD around a relatively low-mass black hole ($\sim10^6~M_\odot$; Puchnarewicz
\etal 1995). However the original {\sl IUE} data were statistically poor,
giving an unreliable flux determination and little information to constrain the
UV continuum slope.

Thus to continue our investigations of the physics and geometry of \rej,  we
have obtained a UV spectrum with the Faint Object Spectrograph (FOS) on the 
{\sl Hubble Space Telescope} ({\sl HST}). This spectrum is a vast improvement
over that obtained with {\sl IUE}, covering a range of $\sim$1200-3000~\AA\ (in
the rest-frame) with a resolution of 1-2~\AA\ FWHM and  signal to noise ratio
of $\sim$15 per diode. In this Letter, we report the first results from this
observation.

\begintable{1} 
\caption{{\bf Table 1.} {\sl HST}-FOS observation log}
\halign{        #\hfil 
     &\quad\hfil#\hfil\quad    
     &\quad\hfil#\hfil\quad   
     &\quad\hfil#\hfil\quad   \cr 
  Grating & resolution & exp time  & range \cr
          & FWHM (\AA) & (seconds) & (\AA) \cr
\noalign{\medskip} 
   G130H & 0.96 & 4420 & 1090-1600 \cr
   G190H & 1.41 & 2460 & 1580-2330 \cr
   G270H & 2.01 &  670 & 2230-3300 \cr
} 
\tabletext{All spectra were taken through the 0.86$^{\prime\prime}$ aperture
and with the {\sc blue} detector.}
\endtable 

\section{Observations and Results}

\rej\ was observed by {\sl HST} on 1997 January 31 using three gratings (G130H,
G190H and G270H) covering the range 1100~\AA\ to 3300~\AA. All spectra were
taken with the {\sc blue} detector; details of the observations, including
exposure times, are given in Table~1. The data were reduced using the standard
{\sl HST} archive pipeline processing.  Errors on the observed flux in each
diode are typically $\sim$10 per cent in the 1100-1500~\AA\ range and 5-10 per
cent between 1500~\AA\ and 3300~\AA. The flux and slope of the {\sl HST}-FOS
spectrum are consistent with the earlier {\sl IUE} data.

Scattered light in the shortest wavelength gratings can be a problem for very
red sources observed with {\sl HST}-FOS (see \eg the {\sl HST}-FOS Instrument
Science Reports 114 and 115). The pipeline corrects for scattered light by
measuring the counts accrued in the unilluminated pixels of the array  (pixels
31-130) after subtraction of the dark particle-induced current, then
subtracting the mean level in this pixel range from the full G130H spectrum.
The wavelength dependence of the scattered light (if any) is poorly understood
however, and in some cases this first order correction is not adequate and
residual scattered light remains. Examination of the raw counts and flux files
for \rej\ shows that the scattered light comprised at most only 20 per cent of
the continuum flux at $\sim$1500~\AA\ before correction by the pipeline, and
appears to have been effectively removed. Thus no further corrections have been
made to these data to account for scattered light.

The G130H/{\sc blue} configuration has shown an increase in sensitivity of
$\sim$5 per cent in the 1250 to 1600~\AA\ range from July 1995 to late December
1996. Therefore the G130H fluxes in this range  were reduced by 5 per cent to
correct for this change in sensitivity. Fluxes in the 1200~\AA\ to 1250~\AA\
range were reduced by a linear ramp from 0 per cent to 5 per cent respectively.

\beginfigure*{1}
\psfig{figure=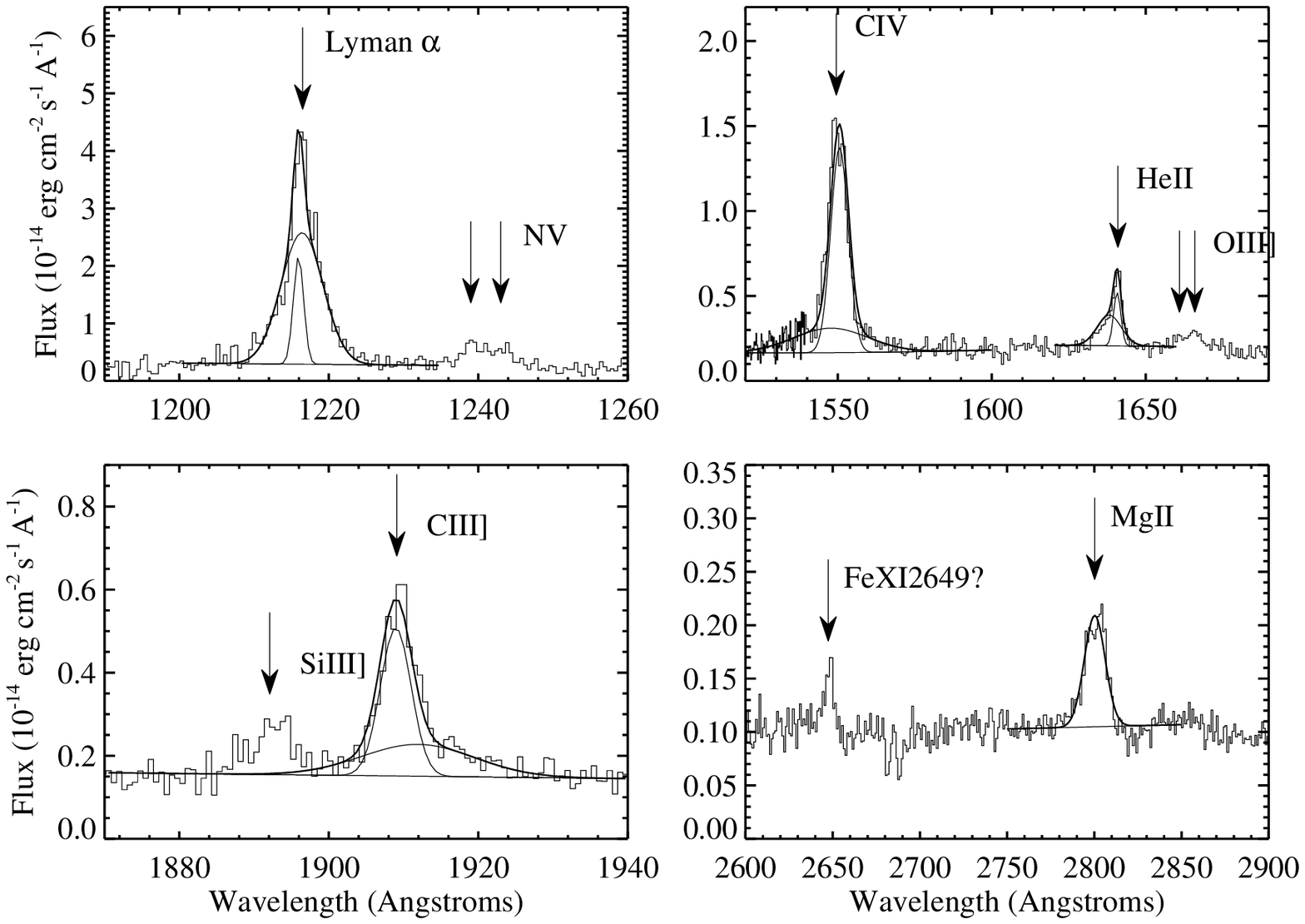,height=5.0in,width=7in,angle=0}
\caption{{\bf Figure 1.} Gaussian fits to emission lines in the
spectrum of \rej. Each portion of the spectrum has been redshifted into the
rest-frame of the AGN (z=0.043) and the expected positions of lines commonly
found in AGN spectra are indicated. The spectra have been binned by a factor of
two for clarity.}
\endfigure

\subsection{UV emission lines}

Rest-frame line positions and full widths at half maximum (FWHM) were measured
from the spectrum for the strongest emission lines and the results are listed
in Table 2. Gaussian profile fitting was used to represent the lines while the
local continuum was modelled as a second-order polynomial. Positions and FWHM
of the lines are the centre and FWHM of the best-fit Gaussian profile
respectively. Errors on the FWHM are difficult to determine due to problems
such as continuum placement, blending with other lines, the use of multiple
components in the fitting procedure and the assumption of a Gaussian profile
for all components. Estimates of the errors have been made nevertheless, and
were derived by comparing the data with the  minima and maxima of `reasonable'
models, taking into account the errors on individual data points. A redshift of
$z$=0.043, measured from the peak positions of the strongest emission lines,
was assumed. Fits to the \lya, \civ, \heii, \ciii\ and \mgii\ lines are shown
in Fig 1; in this figure we also indicate a line at a rest wavelength of
2647.4~\AA\ which has been identified with \fexi. This suggests the presence of
highly-ionized gas in \rej\ and its implications are discussed further in
Section 3.2. 

In most cases (except \mgii), a single Gaussian gave a poor fit to the data, 
so additional Gaussians were used. For Ly$\alpha$\ and \heii, components from
the narrow line region (NLR) were resolved with FWHM of $\sim$500~\kms. NLR
components were not seen in \civ, \ciii\ or \mgii\ however. Although  \civ\ is
dominated by a feature with a FWHM of $\sim$1300~\kms, it appears to have a
much broader  (FWHM$\sim$5500~\kms) underlying component which contains about a
third of the total line flux.  Rodr\'iguez-Pascual, Mas-Hesse \& Santos-Lle\'o
(1997) have also reported broad underlying components in the high-ionization
lines  of several NLS1s from {\sl IUE} spectra, with FWHM between 5000~\kms\
and 10 000~\kms. However, while Rodr\'iguez-Pascual \etal (1997) find these
features underlying \lya, \civ\ and \heii\ in their sample, for \rej, only the
\civ\ line has a significant broad component. There is some evidence of an
additional underlying component in \ciii\ although formally only an upper limit
is obtained (FWHM$<$4300~\kms); the feature  may instead be due to
contamination by other lines or blends (see Fig 1).

\subsubsection{Full widths at half maximum}

The UV emission lines are all  relatively narrow: the broad components of \lya\
and \heii\ have FWHM of less than 2000~\kms, and \mgii\ has a  FWHM of only
1500~\kms.  These FWHM are very low compared to AGN in general, for example,
Brotherton \etal (1994) find mean FWHM for \ciii\ and \mgii\ of 6900~\kms\ and
4200~\kms\ respectively (for their radio-quiet objects). The mean \civ\ FWHM of
the radio-quiet objects from the Wills \etal (1993) sample is 5400~\kms, which
is similar to the very broad underlying feature seen in \rej, but much broader
than the FWHM of the overall profile ($\sim$1400~\kms). 

When measured from optical spectra with a similar velocity resolution, the
broad component  FWHM of the Balmer lines for \rej\ are also narrow  (values of
1500~\kms\ and 1800~\kms\ for \Hb\ and \Ha\ respectively; Puchnarewicz \etal
1995), and similar to those of the UV lines presented here. Thus low velocities
in the line-emitting regions of \rej\  are not confined to the Balmer lines,
but are typical of low-ionization lines in the UV (\eg \mgii) and high
ionization lines (\eg \lya\ and \civ).  

\begintable{2} 
\caption{{\bf Table 2.} UV emission lines}
\halign{   \hfil#\hfil 
          &\quad#\hfil\quad    
     &\quad\hfil#
     &\quad\hfil#\quad   \cr 
 Position & Identification &  FWHM        \cr 
   \AA\   &                &  km s$^{-1}$ \cr 
\noalign{\medskip} 
  1216.4  & \lya\ (broad)                             & 1600$^{+1800}_{-200}$ \cr 
  1216.0  & \lya\ (narrow)                            &  400$^{+700}_{-200}$ \cr 
\noalign{\smallskip} 
  1547.7  & C{\sc IV}$\lambda\lambda$1548,1551 (very broad) & 5500$^{+3000}_{-900}$ \cr 
  1550.6  & C{\sc IV}$\lambda\lambda$1548,1551 (broad)      & 1300$^{+500}_{-500}$ \cr 
\noalign{\smallskip} 
  1638.2  & He{\sc II} (broad)                        &  1700$^{+500}_{-700}$ \cr 
  1640.8  & He{\sc II} (narrow)                       &   500$^{+200}_{-200}$ \cr 
\noalign{\smallskip} 
  1912.0  & C{\sc III}] (very broad)                       &  $<$4300 \cr 
  1909.0  & C{\sc III}] (broad)                      &   800$^{+400}_{-200}$ \cr 
\noalign{\smallskip} 
  2647.4  & [Fe{\sc XI}]$\lambda$2649                &  700$^{+300}_{-300}$ \cr 
\noalign{\smallskip} 
  2800.2  & Mg{\sc II}                                &  1500$^{+400}_{-400}$ \cr 
}
\tabletext{
}        
\endtable 

\subsection{UV continuum slope}

The slope of the continuum was measured by fitting a power-law to the data,
having first removed all emission line features and regions of poor data. The
best-fit was obtained for a slope, $\alpha_{\rm uv}$, of 0.9 (all indices, 
$\alpha$, are defined such that $F_\nu\propto\nu^{-\alpha}$). This is very soft
when compared to the typical UV slopes of quasars where the BBB emission is
strong (\eg Neugebauer \etal (1987) and Francis \etal (1991) find median slopes
of 0.2-0.3 for their quasar samples). 

\subsection{UV to X-ray spectrum}

We have also obtained a quasi-simultaneous measurement of the strength of the
soft X-ray spectrum with the High Resolution Imager (HRI; David \etal 1996)  on
{\sl ROSAT} (Trumper 1983). These data were taken on 1996 November 12 and yield
a count rate of 0.61$\pm$0.02 count sec$^{-1}$. This is consistent with the
earlier {\sl ROSAT}-PSPC  spectrum taken 1991 November (Puchnarewicz \etal
1995).

\beginfigure{2}
\psfig{figure=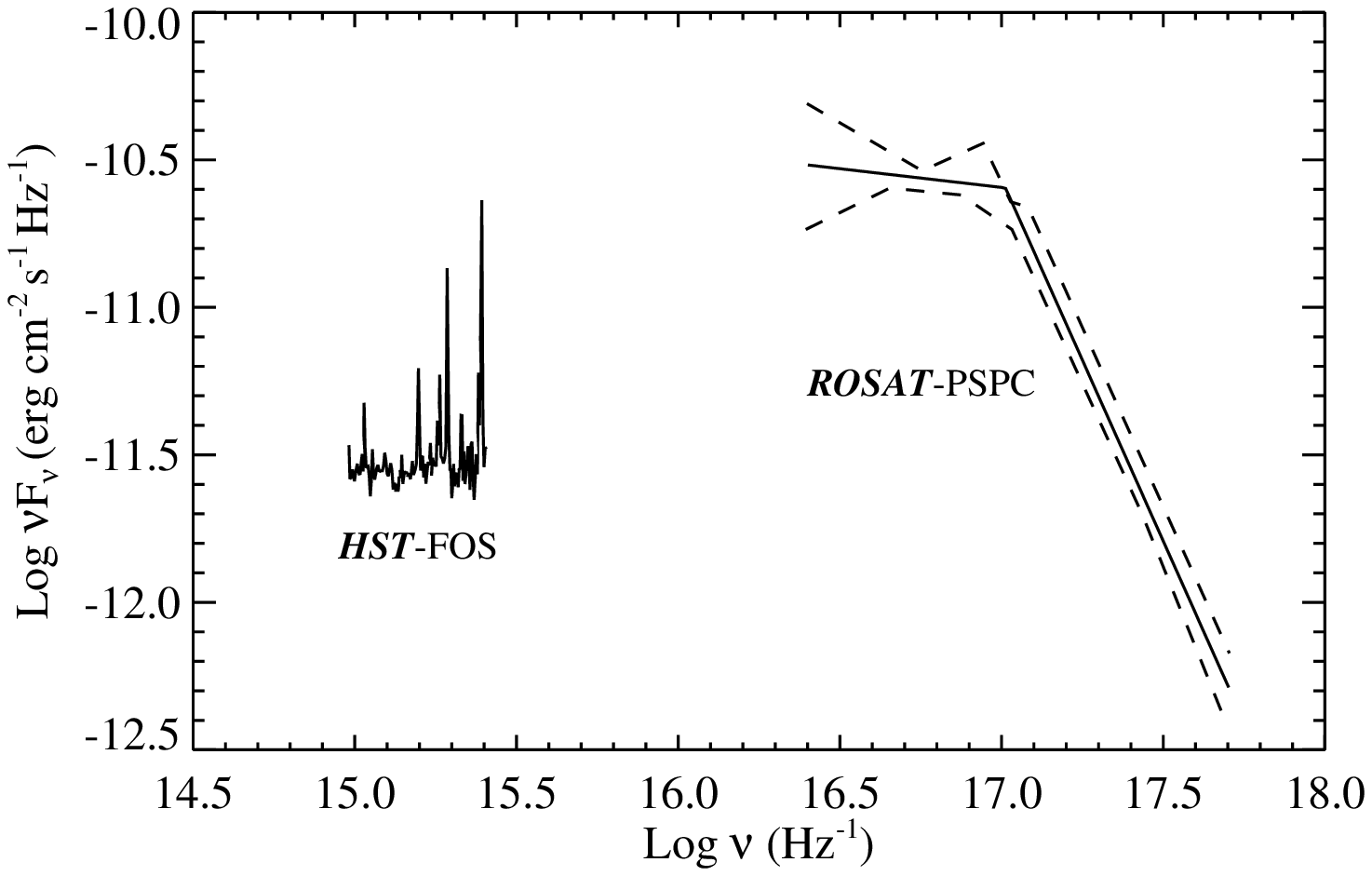,height=2.0in,width=3.0in,angle=0}
\caption{{\bf Figure 2.} The UV to X-ray spectrum of \rej\ (shown as a solid
line), combining the {\sl HST}-FOS and {\sl ROSAT}-PSPC data (normalized to the
HRI flux), and plotted in the rest-frame of the AGN. The dashed lines indicate
90 per cent errors on the broken power-law fit to the {\sl ROSAT}-PSPC data.}
\endfigure

A spectrum of \rej\ from the UV to X-rays (3000~\AA\ to 2~\keV) is shown in
Figure 2. This spectrum combines  the {\sl HST}-FOS data with the model fitted
to the {\sl ROSAT}-PSPC spectrum (from Puchnarewicz \etal 1995) which has been 
normalized to the HRI count rate. Assuming isotropic emission from the UV to
soft X-rays, we measure a luminosity from 3000~\AA\ to 2~\keV\ of
3$\times10^{44}$ erg s$^{-1}$ (the Hubble constant, $H_0$=75 km s$^{-1}$
Mpc$^{-1}$ and deceleration parameter, $q_0$=0; the unobserved 1200~\AA\ to
0.1~\keV\ part of the spectrum was interpolated linearly from the observed
data). The 0.1 to 2~\keV\ spectrum dominates the luminosity overall in this
range, with $L=2\times10^{44}$ erg s$^{-1}$. 

The figure shows the unusually high ratio of soft X-ray to UV flux; the ratio
of the 0.2~\keV\ to 1200~\AA\ fluxes is $\sim$200, whereas for AGN in general,
this ratio is usually {\sl less} than 1. In addition, there is {\sl no} sign
of any steep rise in the UV continuum towards the EUV which is required to meet
the soft X-ray data. It suggests that the relative contribution of the big
bump, even at $\sim$1200~\AA, is very weak. 

Reddening by dust is unlikely to be the cause of the weak UV flux. The UV
continuum doesn't turn down towards the blue as would be expected (\auv=0.9;
Section 2.2) and there is no sign of the broad 2200~\AA\ dust absorption
feature. In the optical, the Balmer decrement is relatively low ($\sim$3;
Puchnarewicz \etal 1995) while the strong soft X-ray spectrum, which shows no
evidence for warm absorber edges, implies that columns of accompanying cold or
warm gas must be low. 

\section{Discussion}

\rej\ is  one of only a handful of AGN which have been selected by the strength
of its EUV emission. It has a very hot BBB (Puchnarewicz \etal 1995) and a very
soft 2-10~\keV\ power-law continuum slope ($\alpha$=1.6; Pounds \etal 1995).
Both its permitted {\sl and} forbidden optical emission lines appear to be
dominated by a line region which is `intermediate' in terms of its velocity,
and perhaps also in its density and position relative to the typical broad and
narrow line regions (Mason \etal 1996). 

With these {\sl HST} data, we have probed further into the emission line
regions of \rej, investigating the production of the high-ionization lines
which, in some systems, are thought to be produced closer to the central black
hole. We have also measured the shape and strength of the UV continuum, and
searched for a rise in the slope which would indicate the emerging low energy
tail of the big bump and constrain accretion disc models. 

\subsection{High-temperature big bump}

The lack of any significant big bump emission in the UV combined with the high
energy turnover in soft X-rays define a high-temperature extreme for current
models of this component. The data presented here ({\sl HST} plus HRI) are
consistent with the earlier {\sl IUE} plus PSPC spectra,  demonstrating a
remarkable degree of stability for \rej, especially when considered with other
objects in the NLS1 class which have shown extreme changes on long and short
timescales. For example, RE~J1237+264 changed by a factor of 70 in 18 months;
Brandt, Pounds \& Fink 1995), while the  soft X-ray flux of IRAS~13224--3809
has been observed to double in only 800 sec (Boller \etal 1993).

The apparent lack of variability in the UV to X-ray spectrum of \rej\  can also
provide a useful discriminant for models of the big bump. Pounds \etal (1995)
have  suggested that \rej\ may be a supermassive analogue of a galactic black
hole candidate (GBHC) in a high state, where the system is accreting close to
its Eddington limit. In this state, the X-ray spectrum has a  steep soft
component which is relatively stable, and a weak power-law tail extending to
higher energies which is highly variable (see \eg Tanaka \& Lewin 1995). \rej\
has a stable ultrasoft component and, although Puchnarewicz \etal (1995) showed
that a sub-Eddington accretion rate is consistent with the observed UV to X-ray
continuum, these AD models are poorly constrained and even super-Eddington
discs cannot be ruled out. \rej\ also has a weak power-law tail like a
high-state GBHC, although there was no evidence of significant flux or spectral
variability during the {\sl ASCA} 1-10~\keV\ observation (Pounds \etal 1995),
which contradicts the GBHC analogy.

\subsection{Emission line regions}

As well as measurements of the UV continuum flux and slope, these {\sl HST}
data have revealed a wealth of emission lines, including \lya, \civ, \heii,
\ciii\ and \mgii\ (see Fig. 1 and Table 2). We also detect \fexi; the \fexiopt\
line, a different transition within the same ionization species, has also been
observed in the optical/IR (Puchnarewicz \etal 1995). These lines indicate the
presence of highly ionized gas in \rej\ and, based on theoretical ratios given
by Penston \etal (1984), the observed \fexi/\fexiopt\  flux ratio of $\sim$0.4
suggests emission from a photoionized gas at a temperature of
$\sim4\times10^4$~K or collisionally-ionized gas with T$\sim10^6$~K. The
observed ratio was calculated using the \fexiopt\ data from Puchnarewicz \etal 
(1992) and an \fexi\ flux of 4$\times10^{-15}$ erg cm$^{-2}$ s$^{-1}$
\AA$^{-1}$ measured from the FOS data.  

The composite UV line profiles are all unusually narrow with FWHM of
$\la$2000~\kms. These are similar to the FWHM of the Balmer lines in \rej\
measured from low-resolution optical spectra. In the case of the Balmer lines,
subsequent high-resolution data showed that the profiles were dominated by an
`intermediate' velocity component (FWHM $\sim$1000~\kms; Mason \etal 1996).
This intermediate component was also seen in the forbidden lines, suggesting
that the region which produced much of the Balmer line and forbidden line
emission was intermediate in terms of its density {\sl and} velocity. Thus it
is possible that this same intermediate velocity component dominates the HILs
as well, implying that one region produces the bulk of the HIL, LIL {\sl and}
the forbidden line emission in \rej. This is in contrast to models of more
typical Seyfert 1s developed from reverberation mapping analyses (\eg  Krolik
\etal 1991), where the HIL region lies close to the black hole and has
relatively high velocity, whereas the LIL region lies $\sim$10-100 light-days
away and has relatively low velocity. 

The high-resolution optical data also revealed the presence of a broad
underlying component (\ie  FWHM of $\sim$2500~\kms), suggesting that a `normal'
Balmer line region does exist in \rej, albeit weak. In these FOS data, again
while a low-velocity component dominates in the HIL profiles,  \civ\ does have
a very broad underlying component (FWHM$\sim$5500~\kms), typical of quasars in
general. This suggests the presence of a `normal' HIL region as well, although
again this is weak relative to the lower-velocity gas. 

The apparent difference in BLR geometry between narrow-line and ordinary
Seyfert 1s may be due to a fine-tuning of the conditions necessary for
efficient line production. Baldwin \etal (1995) suggested that line emission is
strongest where the combination of input ionizing continuum and gas density are
optimized for that particular line species. In the case of \rej\ (and perhaps
all NLS1s), this may be at distances relatively far out from the centre,
because either: {\sl 1.}  there is very little line-emitting gas closer to the
nucleus; or {\sl 2.}  the extreme softness of the ionizing continuum 
tends to produce most of the line emission at greater distances where the
density is lower. 

\section{Summary}

We have presented the UV spectrum of \rej\ in the 1200-3100~\AA\ range, and 
supporting quasi-simultaneous soft X-ray data from the {\sl ROSAT} HRI. These
data confirm the existence of a very hot big bump component in \rej, whose
emission is {\sl not} significant in the UV. The big bump appears to have been
stable over a period of five years, which is remarkable for a NLS1 since many
objects in this class have shown extreme variability in the UV and soft X-rays.
The FWHM of the high ionization lines are narrow ($\la$2000~\kms) and similar
to those of the low ionization lines (including \mgii\ and the Balmer lines).
With the apparent dominance of an intermediate velocity component in both the
Balmer lines and \oiiiab, this suggests that {\sl all} types of line emission
in \rej, \ie HILs, LILs and forbidden lines, may be dominated by the same
region of gas. This geometry contrasts sharply with that derived for normal
Seyfert 1s, where the BLR is highly stratified and the HIL and LIL regions are
separated by $\sim$10-100 light-days. The presence of \fexi\ in the UV spectrum
indicates the presence of highly-ionized gas in the system, which may be
collisionally-ionized or photoionized.

\section*{Acknowledgements}

We thank Martin Ward and the referee, Niel Brandt, for their comments and
advice. This work was supported by {\sl HST} grant GO-06600.01-95A and NASA
contract NAS8-39073.

\bye


\vfil\eject

\onecolumn

\noindent{\bf Abstract}

\bigskip\noindent \rej\ has one of the hottest big blue bumps of any Seyfert 1
(\ktbb$\sim$120~\eV) and thus provides a valuable insight into the physics in
the nuclei of active galaxies. In this paper, we present UV spectroscopy of
\rej, taken with the Faint Object Spectrograph on the {\sl Hubble Space
Telescope}. With a spectral resolution of $\sim$1-2~\AA\ FWHM and a typical
signal-to-noise ratio of $\sim$15 per diode, this is one of the first detailed
UV spectra of an object in the narrow-line Seyfert 1 (NLS1) class. The spectrum
probes the physics and kinematics of the high-ionization and coronal line gas,
and the strength and form of the big blue bump component in the UV.

\bigskip\noindent We detect many emission lines, including \lya, \civ,
\heiifull, \ciiifull\ and \mgiifull. We also identify a feature at 2647~\AA\
(in the rest-frame) with highly-ionized iron (\fexi); a line of the same
species (\fexiopt) has also been seen in the optical spectrum. The velocity
widths of the UV lines are relatively narrow (FWHM$<$2000~\kms) although \civ\
appears to have a broad underlying component with a FWHM typical of quasars
($\sim$5500~\kms). The FWHM are similar to those of the optical lines, which
suggests that {\sl all} line emission in \rej, \ie including that of high- and
low-ionization species {\sl and} the forbidden lines, may be dominated by an
intermediate-velocity (FWHM$\sim$1000~\kms), intermediate-density (log $N_{\rm
e}\sim7.5$ cm$^{-3}$) region of gas. The slope of the UV continuum
(\auv$\sim$0.9) is soft (\ie red) relative to quasars and the UV-to-soft X-ray
flux ratio is unusually low (the 0.2~\keV/1200~\AA\ flux ratio is 1/200),
implying that the big blue bump component is very weak in the UV. The present
epoch UV to soft X-ray continuum is consistent with earlier data, demonstrating
that this extreme big blue bump component is also very stable, unlike many
other NLS1s which show extreme patterns of variability.


\vfil\eject

\noindent{\bf Figure 1.} Gaussian fits to emission lines in the spectrum of
\rej. Each portion of the spectrum has been redshifted into the rest-frame of
the AGN (z=0.043) and the expected positions of lines commonly found in AGN
spectra are indicated. The spectra have been binned by a factor of two for
clarity.

\bigskip\noindent{\bf Figure 2.} The UV to X-ray spectrum of \rej\ (shown as a solid line),
combining the {\sl HST}-FOS and {\sl ROSAT}-PSPC data (normalized to the HRI
flux), and plotted in the rest-frame of the AGN. The dashed lines indicate 90
per cent errors on the broken power-law fit to the {\sl ROSAT}-PSPC data.
       
\bye